\documentclass[twocolumn,preprintnumbers,nofootinbib,amsmath,amssymb,showkeys]{revtex4}
\usepackage{graphicx}		
\usepackage{dcolumn}			
\usepackage{bm}				

\begin{document}
\title{Exhaustive exploration of Prisoner's Dilemma parameter space in one-dimensional cellular automata}

\author{Marcelo Alves Pereira$^{\dag}$, Alexandre Souto Martinez$^{\ddag}$}
\affiliation{Departamento de F\'{\i}sica e Matem\'{a}tica - FFCLRP - USP \\
Av. Bandeirantes, 3900, 14040-901 \\
Ribeir\~{a}o Preto, SP, Brazil \\
$^{\dag}$ marceloapereira@usp.br \\
$^{\ddag}$ asmartinez@usp.br}

\author{Aquino Lauri Esp\'{i}ndola}
\affiliation{Departamento de Medicina Social - FMRP - USP \\
Departamento de F\'{\i}sica e Matem\'{a}tica - FFCLRP - USP \\
Av. Bandeirantes, 3900, 14040-901 \\
Ribeir\~{a}o Preto, SP, Brazil \\
aquinoespindola@usp.br}

\date{\today}

\begin{abstract}
The Prisoner's Dilemma (PD) is one of the most popular games of the Game Theory due to the emergence of cooperation among competitive rational players.
In this paper, we present the PD played in cells of one-dimension cellular automata, where the number of possible neighbors that each cell interacts, $z$, can vary.
This makes possible to retrieve results obtained previously in regular lattices.
Exhaustive exploration of the parameters space is presented.
We show that the final state of the system is governed mainly by the number of neighbors $z$ and there is a drastic difference if it is even or odd.
\end{abstract}


\maketitle

\section{Introduction} \label{introduction}
Due to the emergence of cooperation between competitive rational players \cite{stauffer2004,bouchaud2002,anteneodo2002,turner1999}, the \emph{Prisoner's Dilemma} (PD) \cite{dresher,poundstone1992} is one of the most popular games of the Game Theory \cite{neumann1947}.
When it is played repeatedly, one has the Iterated Prisoner's Dilemma \cite{axelrod1981,axelrod1984}.
If the PD is played in a group of players with spatial structure, this version is known as Spatial Prisoner's Dilemma (SPD) \cite{nowak1992}.
These spatial structures may generate chaotically changing spatio-temporal patterns.
Cooperators and defectors coexist, and cooperator proportion oscillates indefinitely.
This occurs when each player interacts with the nearest neighbors, for instance, in a square lattice.
Moreover, adding the interaction with the next nearest neighbors (corresponding to the chess king possible moves) the spatial patterns are smoother.
During the game, cooperators and defectors organize themselves in clusters.
The most interesting dynamics occurs on the borders of these clusters, causing the oscillating behavior of the proportion of cooperators.

The final proportion of cooperators and defectors in the chaotic phase depends on the initial configuration and the magnitude of the parameter $T$ (temptation).
Moreover, the connectivity among players also plays an important role in the dynamics of the clusters \cite{duran2005}.
Studies about PD had been carried out in different topologies such as square lattice \cite{nowak1992}, graphs \cite{nowak2005} and also in complex networks as random graphs \cite{duran2005}, scale-free networks \cite{wu2007}, small-word networks \cite{abramson2001}.
We have used the simplest lattice topology, i.e. one-dimensional lattice to represent regular lattices at any dimensionality \cite{marcelo2007}.
The computational implementation of PD in the one-dimensional case is simpler than in other topologies, and it requires less computational time to run the numerical codes.
In one-dimensional cellular automata, it is simpler to understand the way that oscillations in the cooperator proportion take place \cite{marcelo2007}.
Beyond the topologies, it is also possible to consider the mobility of players \cite{arenzon2007}.

In this paper we present an exhaustive exploration of the parameter space for the IPD in the one-dimensional cellular automata with a variable number of interacting neighbors.
After introducing the model in Section \ref{themodel}, we show the results in Section \ref{resultados}.
Final remarks are presented in the Section \ref{conclusao}.

\section{The Model} \label{themodel}
Consider a cellular automaton in a one-dimensional lattice, with $L$ cells, where each cell represents one player, who has two possible states: $\theta = 1$ ($\theta = 0$) for cooperator (defector).
The automaton has no empty cells, so the cooperator proportions, $\rho_c(t)$, and defectors, $\rho_d(t)$, leads to $\rho_c(t) + \rho_d(t) = 1$.
The initial proportion of cooperators, $\rho_c(0) = \rho_0$, $0 \leq \rho_0 \leq 1$, is an important parameter.
The state of $L\rho_0$ players, which are chosen randomly by a uniform distribution, are set as cooperators and the remaining ones are set as defectors.
The neighborhood of the $i$-th player is defined by $z = (1, 2, \ldots, L)$.
If $z$ is even, there are $\alpha = z/2$ adjacent interacting players to the right and to the left hand side of this player.
If $z$ is odd, each side has $\alpha = (z-1)/2$ players and player $i$ interacts with his/her own state (self-interaction) \cite{marcelo2007,ricardo2006,alves2000}.
In addition to $\rho_0$ and $z$ the other free parameter in this model is the temptation $T$ in the conflict range $1 \leq T \leq 2$.

Consider two players $i$ and $j$ playing the PD.
The payoff of player $i$ due to interaction with player $j$ is given by
$g_{\theta_i,\theta_j} = \theta_i \theta_j + T(1-\theta_i \theta_j)\theta_j$,
where $\theta_j$ is the state of player $j$, with $j = (1, 2, \ldots, L)$.
The total payoff, $P_i$, of player $i$ is:
$P_i = \sum_{{\cal V}_i} g_{\theta_i,\theta_j}$, , where ${\cal V}_i$ is the neighborhood of the $i$-th agent.
Since the payoff of each player depends on $z$, the macroscopic regime, $\rho_c$, also depends on it.
Player $i$ will compare $P_i$ to $P_k$, where $P_k$ is the payoff of $k = (1, 2, \ldots, z)$ set of players.
If $P_i < P_k$, player $i$ copies the state of the player with the highest payoff, otherwise player $i$ does not change his/her current state.
The dynamics of the model is totally deterministic.
This strategy of copying the state of the neighbor that had the highest payoff is the Darwinian Evolutionary Strategy.
Others evolutionary strategies can be adopted, like the Pavlovian one \cite{fort2005}.
The states of the players are updated synchronously and they play until the system reaches a stationary or dynamical equilibrium regime.

The cooperator proportion, $\rho_c(t,T,\rho_0,z)$, depends on time, temptation, initial proportion of cooperators, and the number of interacting players.
The dependence of $\rho_c$ as a function of $\rho_0$ and $z$ is commonly neglected, possibly due to the fixed lattice restriction in a $d$-dimensional space.

The asymptotic cooperator proportion, $\rho_\infty(T,\rho_0,z)$, is obtained when the system reaches the steady state, which represents the final phase for the set of parameters $(T,\rho_0,z)$.
The dependence of $\rho_\infty$ on $z$ can be understood due to the number of interacting cooperators, $c$, with $0 \leq c \leq z$, in the neighborhood of each player.
When player $i$ interacts with $c_i$ cooperators out of $z$ neighbors, his/her payoff is \cite{duran2005,ricardo2006}:
$P_{i}^{(c_i)}(\theta_i) = [T-(T-1)\theta_i]c_i$.
Some useful relations follow immediately: for a cooperator $P_{i}^{(c_i)}(1) = c$, while for a defector $P_{i}^{(c_i)}(0) = cT$.
For $T > 1$, $P_{i}^{(c_i)}(0) > P_{i}^{(c_i)}(1)$ and $P_{i}^{(c)}(\theta) \geq P_{i}^{(c - 1)}(\theta)$.
Transitions in $\rho_\infty(T)$ occur when temptation crosses threshold values.
In the conflict range, $1 < T < 2$, these transitions are controlled by \cite{duran2005}:
$T_c(n,m) = (z-n)/(z-n-m)$,
where $0 \leq n < z$ and $1 \leq m \leq \mbox{int}[(z-n-1)/2)]$ are integers.

\section{Results}\label{resultados}
We have used a one-dimensional cellular automaton with $L = 1,000$ cells, with $L \rho_0$ cells set as cooperators and the remaining ones as defectors.
The asymptotic cooperator proportion, $\rho_\infty$, is obtained from the mean values of an ensemble of 1,000 configurations for the same initial parameters.
The parameter $T$ increases in steps $\Delta T = 0.01$ in the range $1 < T < 2$ and $\rho_0$ increases in steps $\Delta \rho_0 = 0.1$ and the intermediate values are linearly interpolated.

It could seem meaningless to consider $T = 1.00$, as the cooperators and defectors have the same payoff, when one plays against the other.
However, the total payoff of each player depends on the neighborhood, then, if the player belongs to a cooperative cluster he/she has a higher payoff than the player from a defective one.
In the cooperative/defective clusters border, the differences among payoffs are essential to determine the system dynamics \cite{marcelo2007}.

Results for $\rho_0 = 0$ and $\rho_0 = 1$ are the trivial cases due to the Darwinian Evolutionary Strategy.
In a population of cooperators (defectors) it is not possible to emerge a defector (cooperator), because the players only can copy the states of their neighbors.
Mutations are not allowed in our model, i.e. the noise of the system is null \cite{mukherji1996}.

Our results are equivalent to those obtained in the square lattices, which are briefly reviewed in the following.
Consider four scenarios.
First, defectors can dominate the system and determine the complete extinction of cooperators, leading the system to a defective phase ($\rho_\infty = 0$).
Second, defectors can increase and domain the system, but cooperators are not extinguished, resulting in a defective phase as well ($0 < \rho_\infty < 0.5$).
Third, cooperators may domain the system forming a cooperative phase ($0.5 < \rho_\infty < 1.0$).
And finally, cooperators can extinguish the defectors determining a cooperative phase ($\rho_\infty = 1.0$).

\begin{figure}[!htb]
(a) \\
\includegraphics[width=1.0\linewidth]{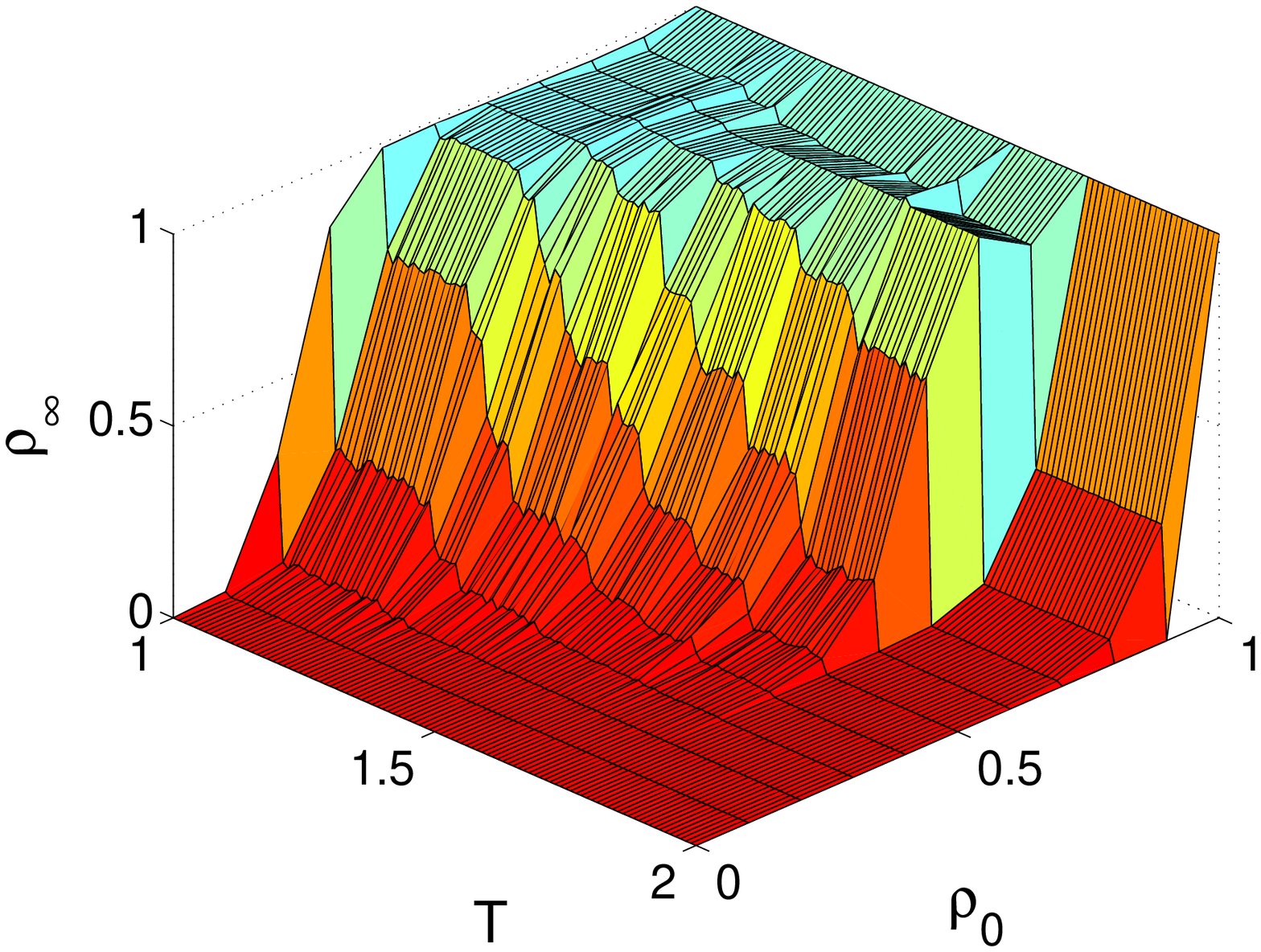} \\
(b) \\
\includegraphics[width=1.0\linewidth]{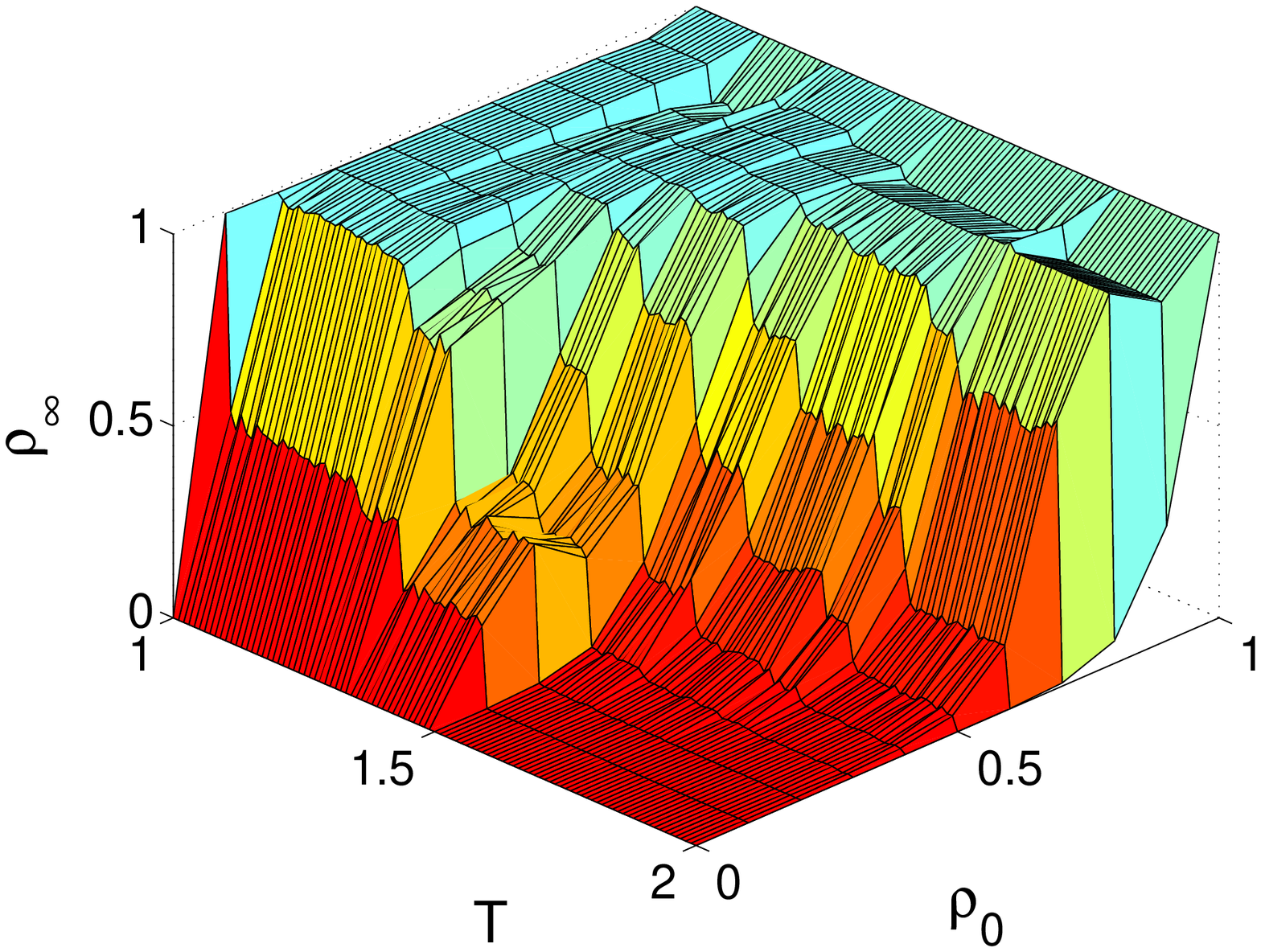}
\caption{
Phase diagram, $\rho_\infty$ (asymptotic proportion of cooperators) as function of $T$ (temptation), $\rho_0$ (initial proportion of cooperators) and $z$ (number of interacting players of each player) plotted as a surface.
(a) $z = 8$ (without self-interaction);
(b) $z = 9$ (with self-interaction).
\label{fig_superficies}}
\end{figure}

Figures \ref{fig_superficies}a and \ref{fig_superficies}b show the surface of $\rho_\infty$ plotted as a function of $T$ and $\rho_0$, for $z = 8$ (without self-interaction) and $z = 9$ (with self-interaction), respectively.
Differences between the presence/absence of self-interaction are clear.
The region of low values of $T$ and high values of $\rho_0$ is a cooperative phase.
The region of high values of $T$ and low values of $\rho_0$ is a defective phase.
The other two regions, low $T$ and low $\rho_0$ or high $T$ and high $\rho_0$, the value of $\rho_\infty$ is different and depends strongly whether $z$ is even or odd.
The valleys for $\rho_0 \sim 0.9$ are due to the system dynamics.
For higher values of $\rho_0$, the defective clusters formed are tiny in comparison to the cooperative ones.
These defectors exploit theirs cooperator neighbors, but they neither do not invade the cooperative cluster nor are extinguished by the cooperative neighborhood during time evolution.

\begin{figure}[!htb]
\includegraphics[width=0.45\linewidth]{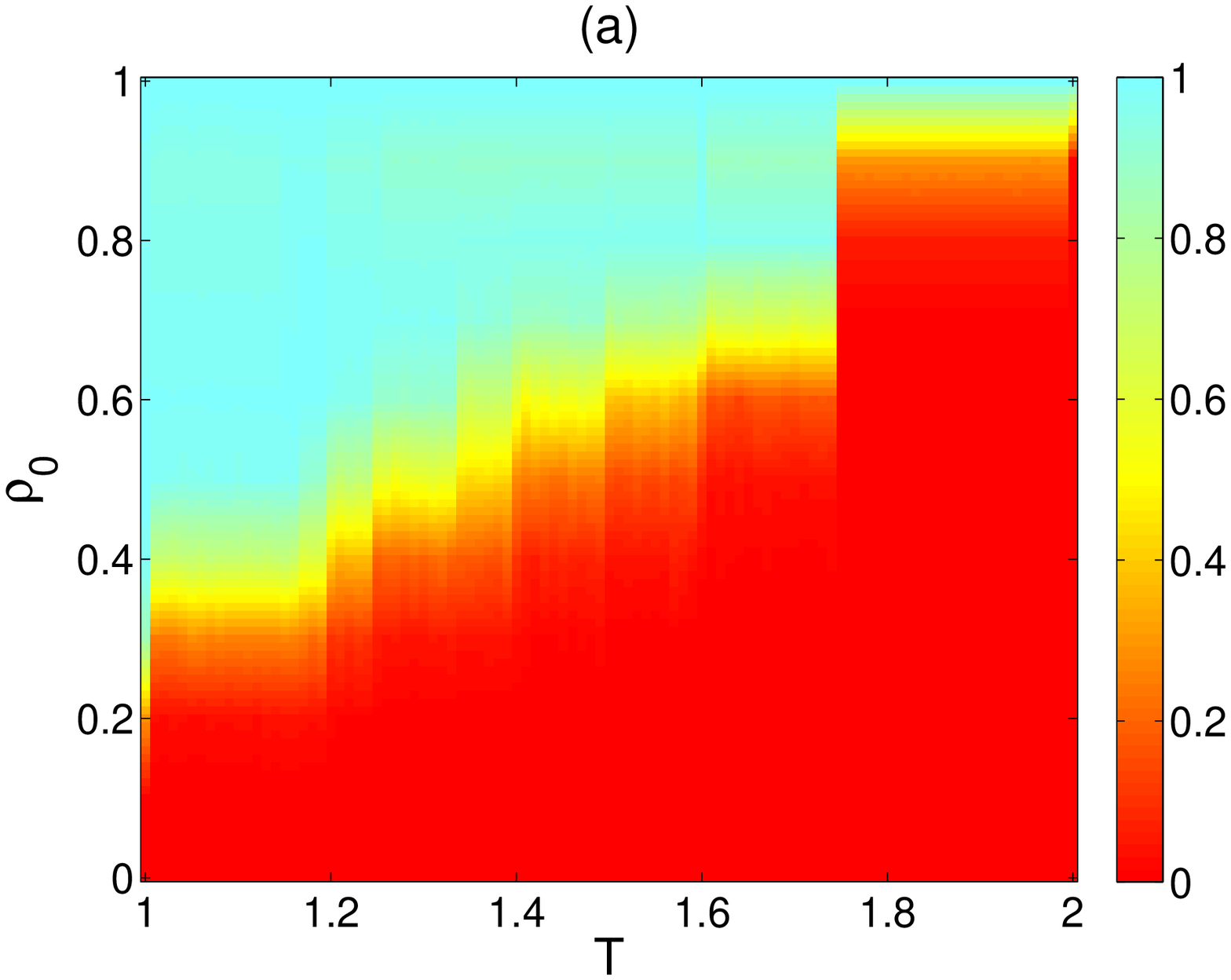}
\includegraphics[width=0.45\linewidth]{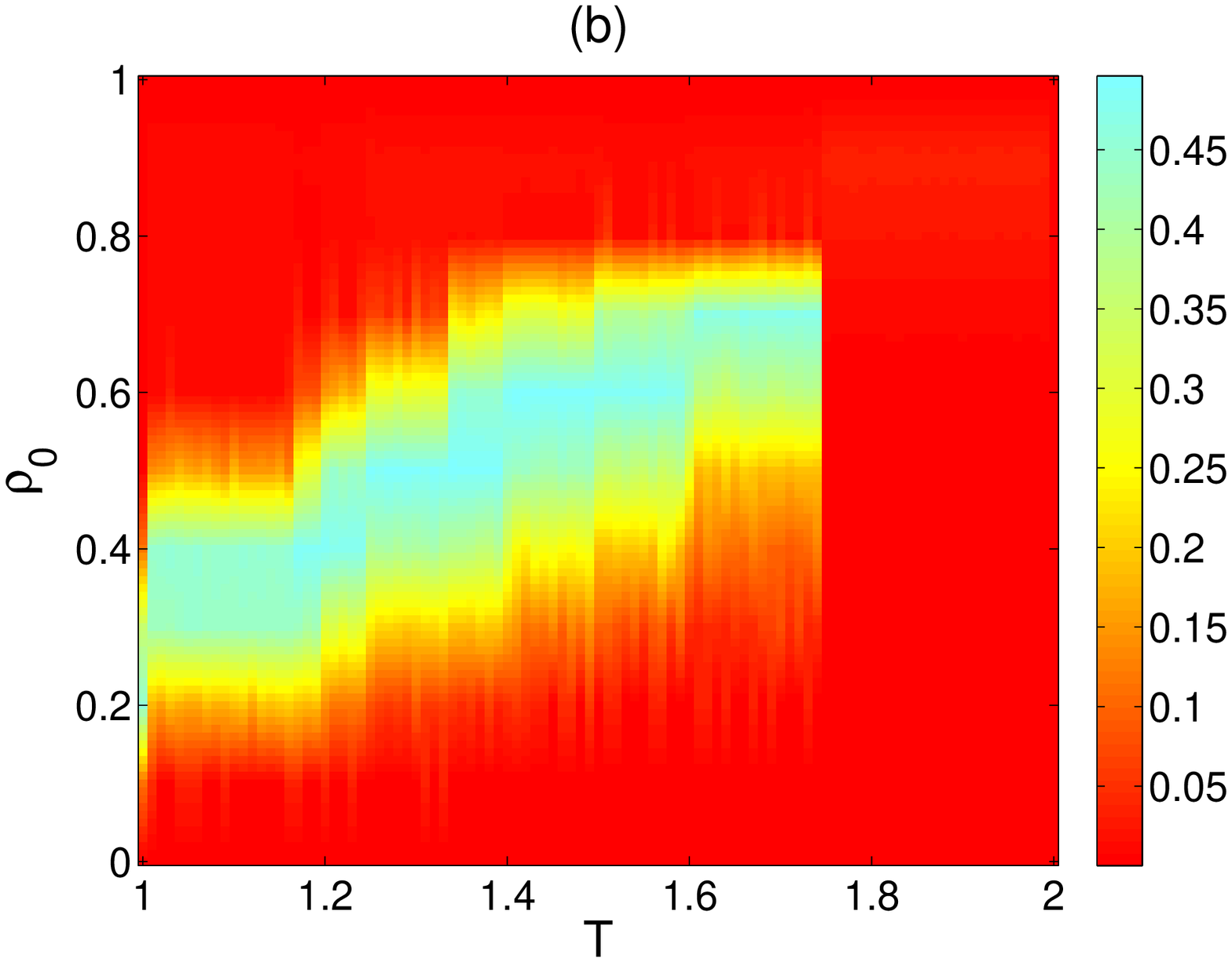} \\
\includegraphics[width=0.45\linewidth]{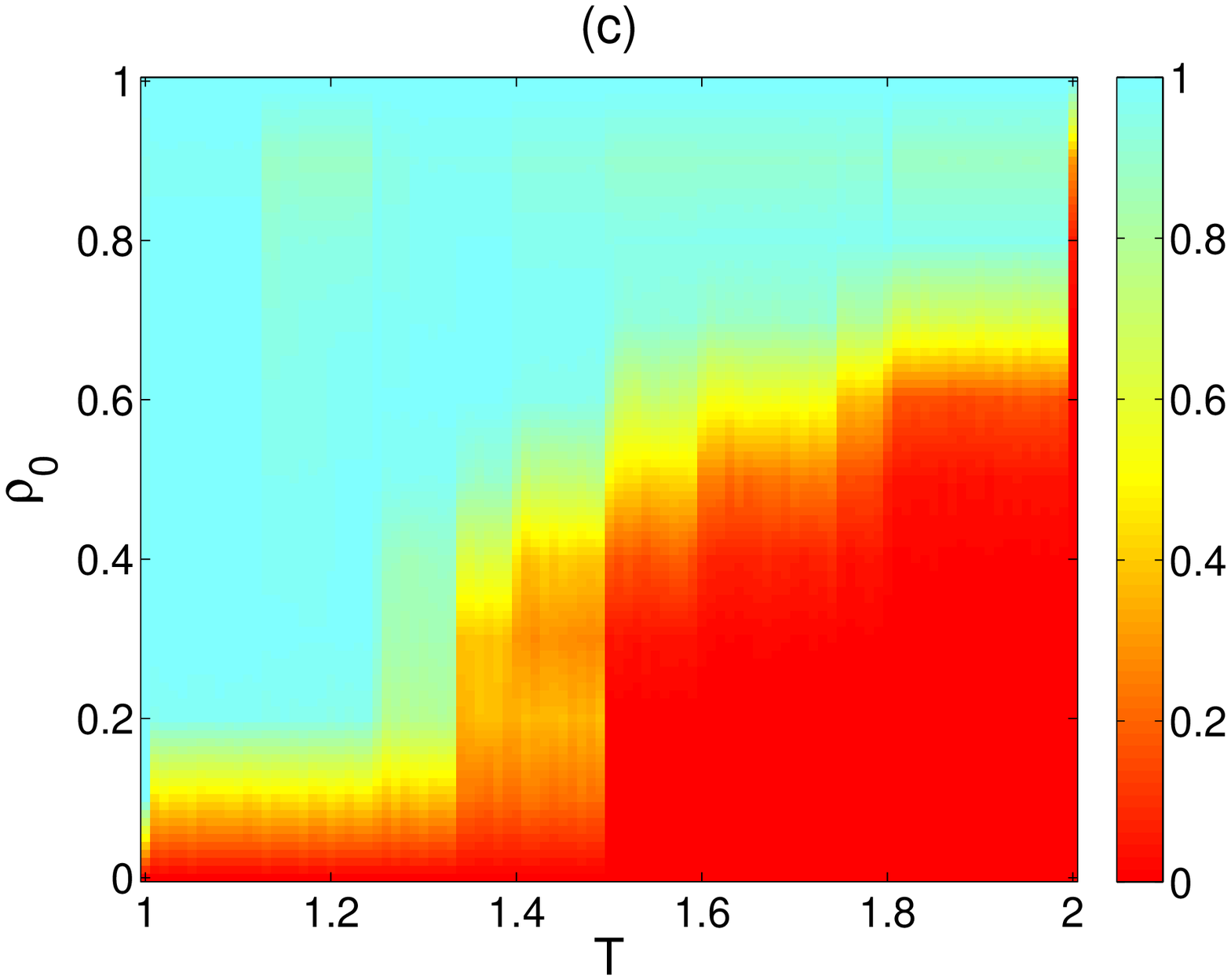}
\includegraphics[width=0.45\linewidth]{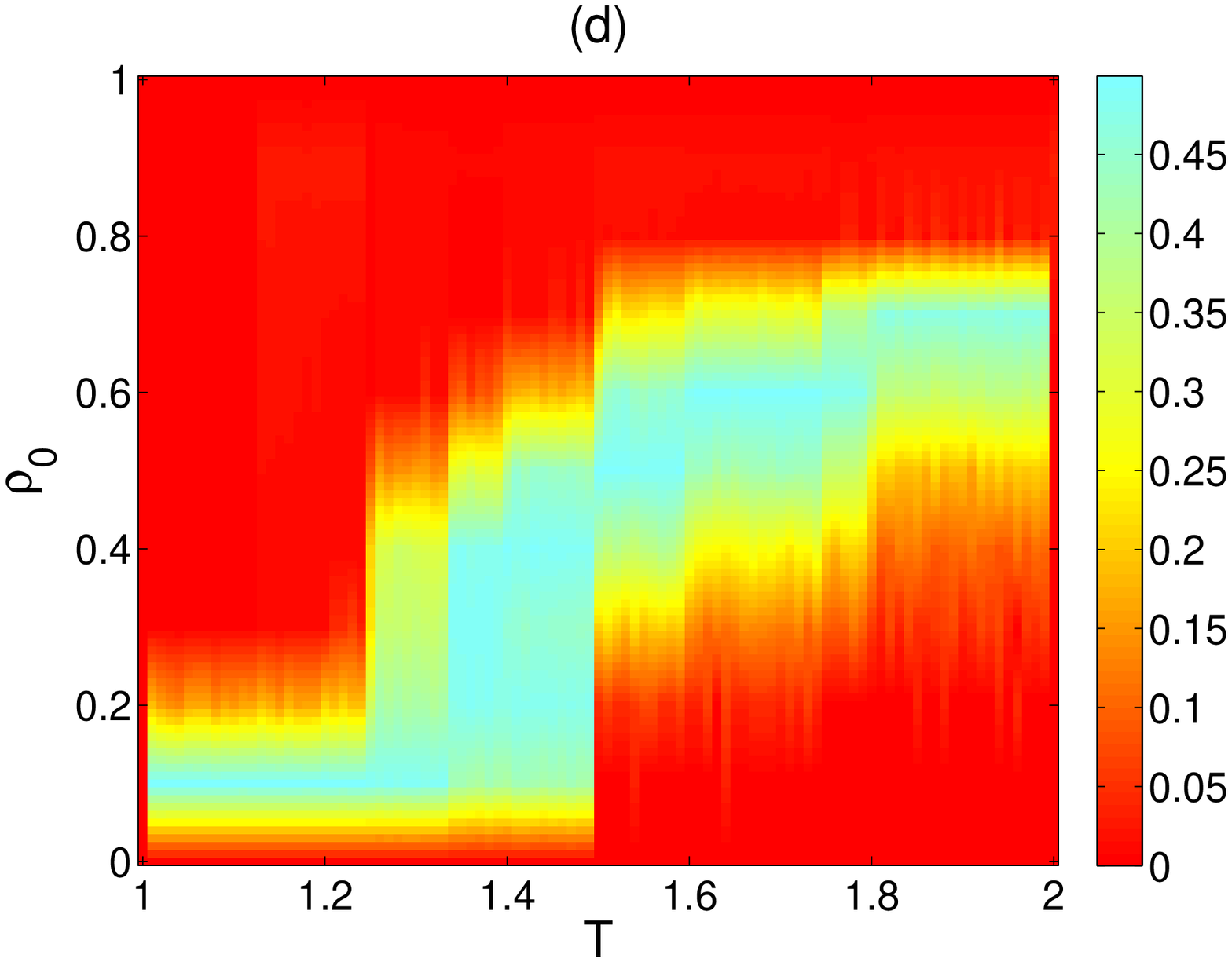}
\caption{
(a) Top view of Figure \ref{fig_superficies}a, $\rho_\infty$ as function of $T$ and $\rho_0$ for $z = 8$ (without self-interaction);
(b) standard deviation of $\rho_\infty$ as function of $T$ and $\rho_0$ for $z = 8$ (without self-interaction);
(c) Top view of Figure \ref{fig_superficies}b, $\rho_\infty$ as function of $T$ and $\rho_0$ for $z = 9$ (with self-interaction);
(d) standard deviation of $\rho_\infty$ as function of $T$ and $\rho_0$ for $z = 9$ (with self-interaction).
\label{fig_contornos}}
\end{figure}

Another visualization of $\rho_\infty$ for $z = 8$ is given in Figures \ref{fig_contornos}a and \ref{fig_contornos}b, and for $z = 9$ in Figures \ref{fig_contornos}c and \ref{fig_contornos}d.
It is equivalent to observe the phase diagram plotted as surface in Figures \ref{fig_superficies}a and \ref{fig_superficies}b from the top view.
The images \ref{fig_contornos}b and \ref{fig_contornos}d are the standard deviation of $\rho_\infty$ due to statistics to avoid the initial configuration dependence.
Figures \ref{fig_contornos}b and \ref{fig_contornos}d, show very high values of standard deviation.
In these regions, small changes in the initial configuration drastically modify $\rho_\infty$ from a cooperative phase, $\rho_\infty > 0.5$, to a defective phase, $\rho_\infty < 0.5$.
Thus, in this region, it is not possible to define the system as cooperative or defective, and this region is considered as the coexistence of cooperative/defective phases.
In other words, the chaotic phase.
The inclusion of self-interaction implies in larger cooperation area in the phase diagram as shown in Figures \ref{fig_contornos}a and \ref{fig_contornos}c.
This means that cooperation prevails when self-interaction is included.
In Figure \ref{fig_contornos}a, $\rho_\infty$ drops abruptly for $T > 1.7$, this rapid decay does not occur in Figure \ref{fig_contornos}c, because self-interaction shifts $T_c$ to higher values.
In Figure \ref{fig_contornos}d, the higher values of standard deviation fulfill a larger area, especially for $T > 1.7$, for the same reasons.
When $T_c$ is shifted, a region that should be cooperative, when the self-interaction is present, becomes defective in the absence of the self-interaction.

\begin{figure}[!htb]
(a) \\
\includegraphics[width=1.0\linewidth]{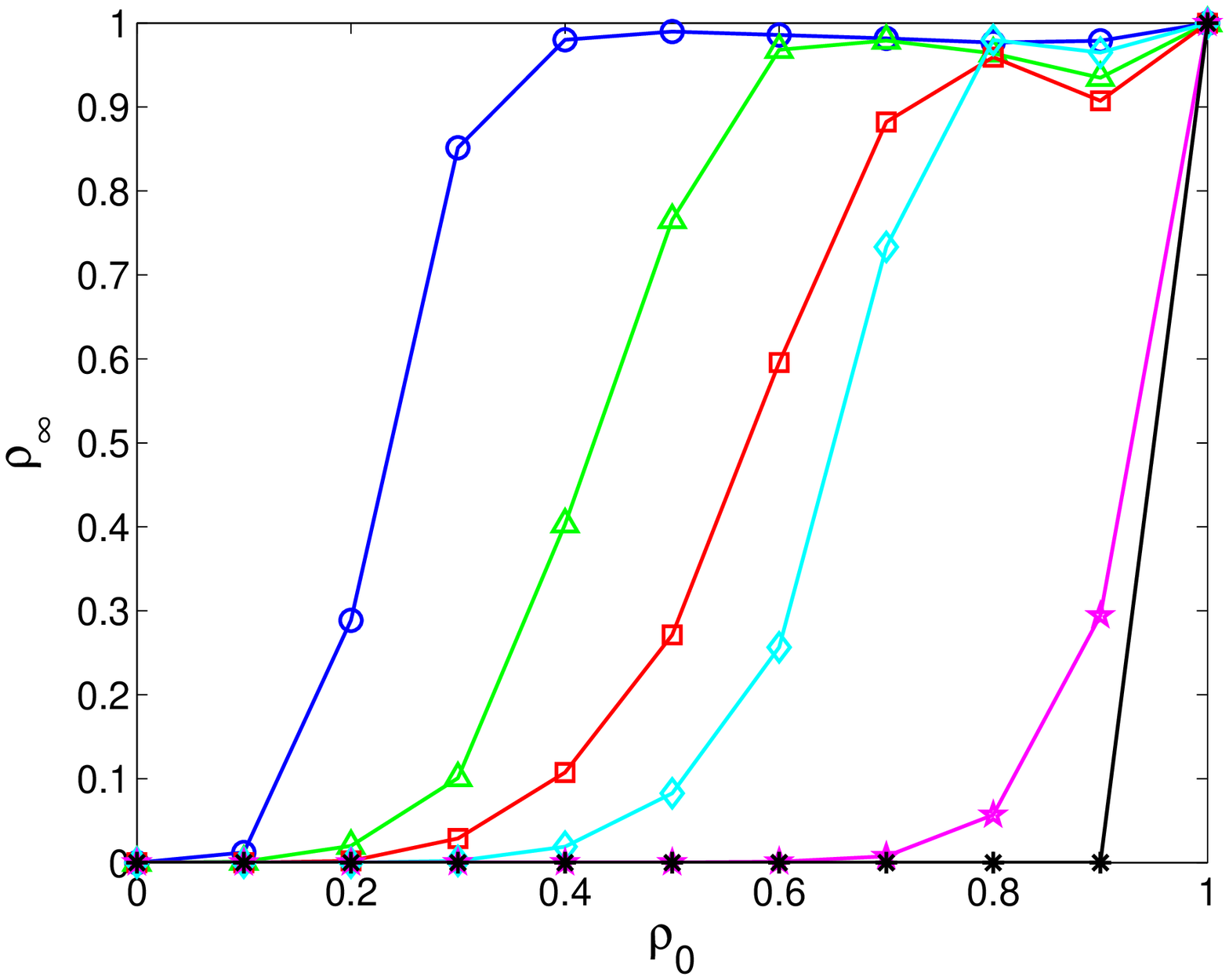} \\
(b) \\
\includegraphics[width=1.0\linewidth]{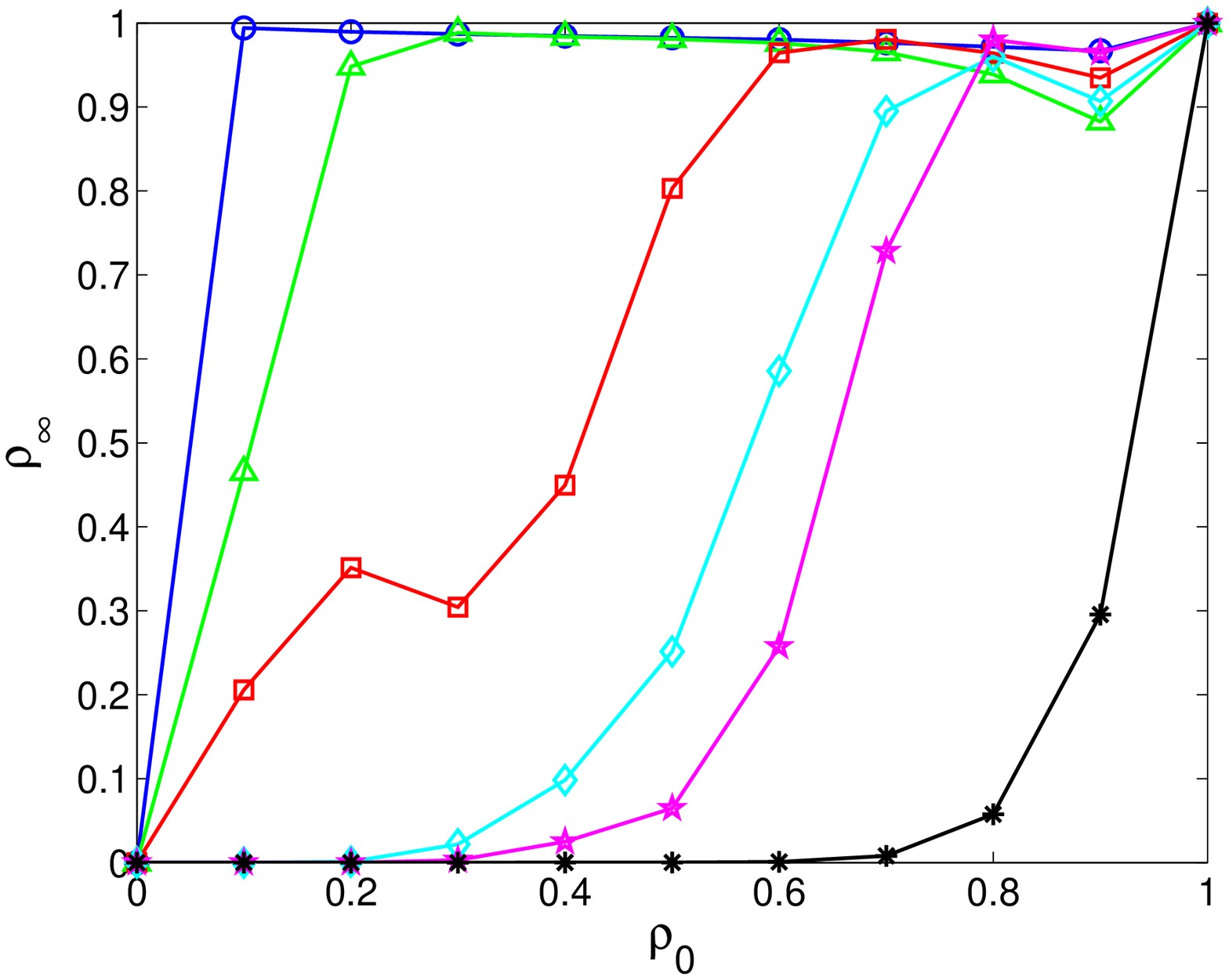}
\caption{
Slice of the plane $\rho_\infty(T,\rho_0,z)\rho_0$ of the surfaces.
Asymptotic proportion of cooperators ($\rho_\infty$) as function of initial proportion of cooperators ($\rho_0$) for (a) $z = 8$ (without self-interaction) and (b) $z = 9$ (with self-interaction).
Asterisks: $2.0$; Stars: $T = 1.8$; Diamonds: $T = 1.6$; Squares: $T = 1.4$; Triangles: $T = 1.2$; Circles: $T = 1.0$.
\label{fig_plots_p0_Ts}}
\end{figure}

The slice $\rho_\infty \rho_0$ of Figure \ref{fig_superficies} shows $\rho_\infty$ as a function of $\rho_0$.
The curves are plotted for $T = (2.0, 1.8, 1.6, 1.4, 1.2, 1.0)$, in Figures \ref{fig_plots_p0_Ts}a for $z = 8$, and Figure \ref{fig_plots_p0_Ts}b for $z = 9$.
The value of $\rho_\infty$ increases in presence of self-interaction in the region $0 < \rho_0 < 0.4$, for $1.0 < T < 1.4$.
Self-interaction also shifts the emergence of cooperation to lower values of $\rho_0$, when compared to a system without self-interaction.
In Table \ref{tab_deslocamento}, one sees the values of $\rho_0$, where $\rho_\infty > 0.5$ occurs for the first time for different values of $T$.
Notice the strong difference concerning the parity of $z$.

The $\rho_\infty$ non-monotonous behavior for intermediate values of $\rho_0$ presented in Figures \ref{fig_superficies} and \ref{fig_plots_p0_Ts}, in the region $1.3 < T < 1.5$ and $0 < \rho_0 < 0.5$  are due to the coexistence phases.
In this region, the standard deviation of $\rho_\infty$ is higher than in the remaining regions.

\begin{table}[tbp]
\centering
\begin{tabular}{c|c|c}
\hline
$T$ & $z = 8$ & $z = 9$ \\ \hline
1.0 & 0.3 & 0.1 \\
1.2 & 0.5 & 0.2 \\
1.4 & 0.6 & 0.5 \\
1.6 & 0.7 & 0.6 \\
1.8 & 1.0 & 0.7 \\
2.0 & 1.0 & 1.0 \\ \hline
\end{tabular}
\caption{Values of $\rho_0$, when occurs $\rho_\infty > 0.5$ for the first time, for different values of $T$, for a system with $z = 8$ (without self-interaction) and $z = 9$ (with self-interaction).}
\label{tab_deslocamento}
\end{table}

\begin{figure}[!htb]
(a) \\
\includegraphics[width=1.0\linewidth]{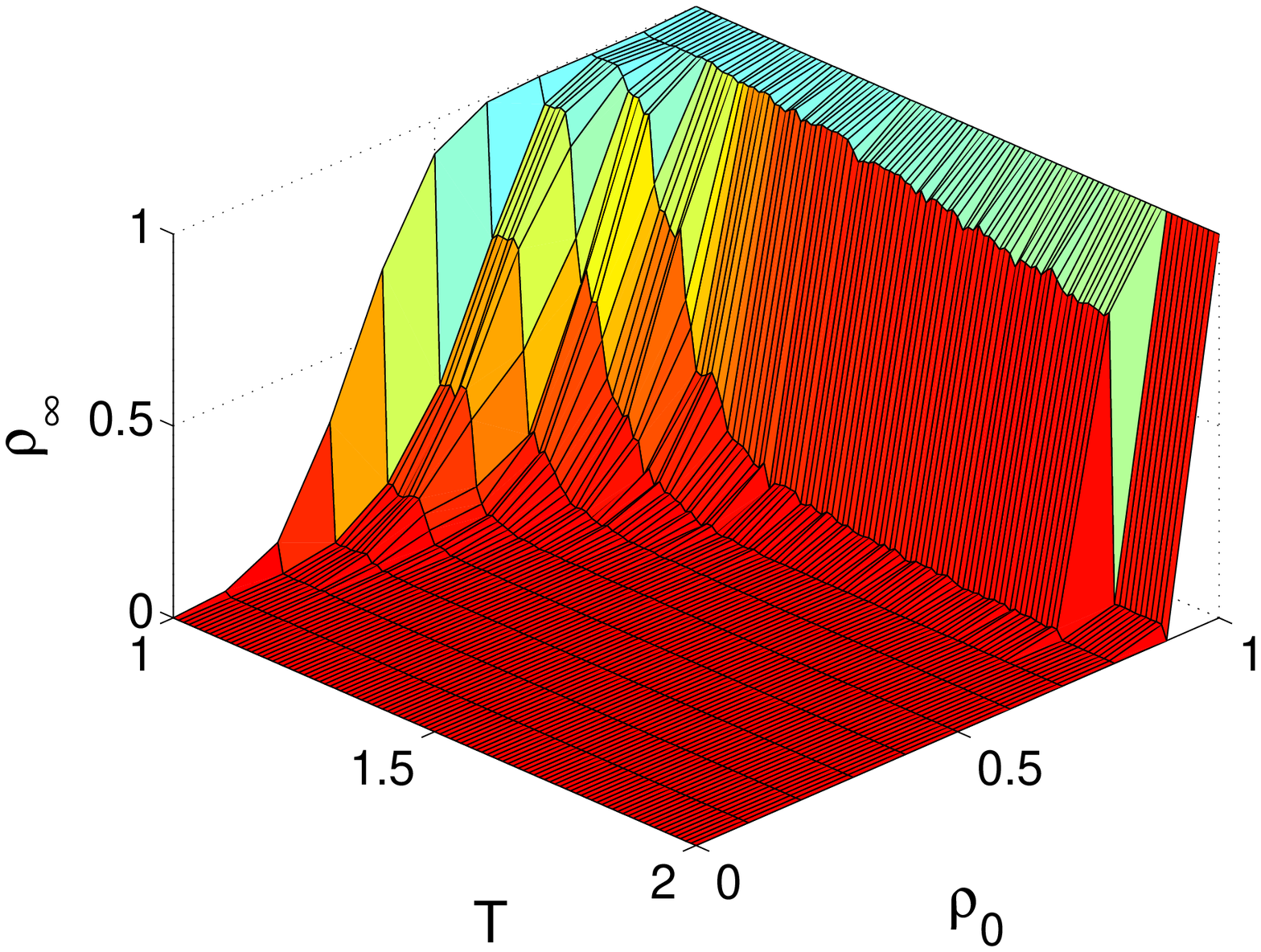} \\
(b) \\
\includegraphics[width=1.0\linewidth]{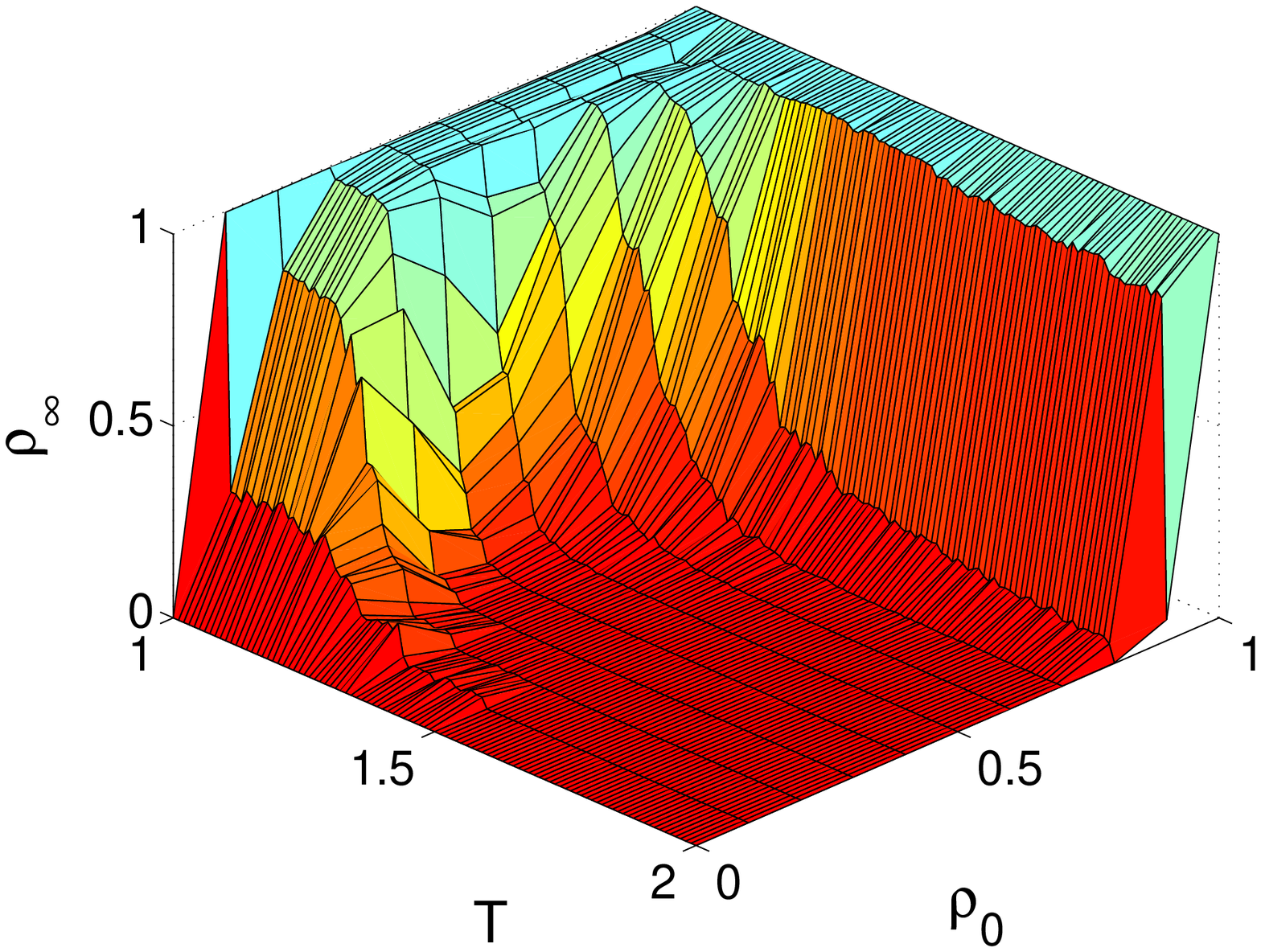}
\caption{
Phase diagram, $\rho_\infty$ (asymptotic proportion of cooperators) as function of $T$ (temptation), $\rho_0$ (initial proportion of cooperators) and $z$ (number of interacting players of each player) plotted as a surface.
(a) $z = 20$ (without self-interaction);
(b) $z = 19$ (with self-interaction).
\label{fig_superficies_converg}}
\end{figure}

To observe the behavior of $\rho_\infty$, when $z$ increases, see the surfaces of $\rho_\infty$ for $z = 20$, in Figures \ref{fig_superficies_converg}a, and $z = 19$, Figure \ref{fig_superficies_converg}b.
Comparing Figures \ref{fig_superficies}a and \ref{fig_superficies}b, one observes that if $z$ is increased, the surfaces become more similar.

\begin{figure}[!htb]
\includegraphics[width=0.45\linewidth]{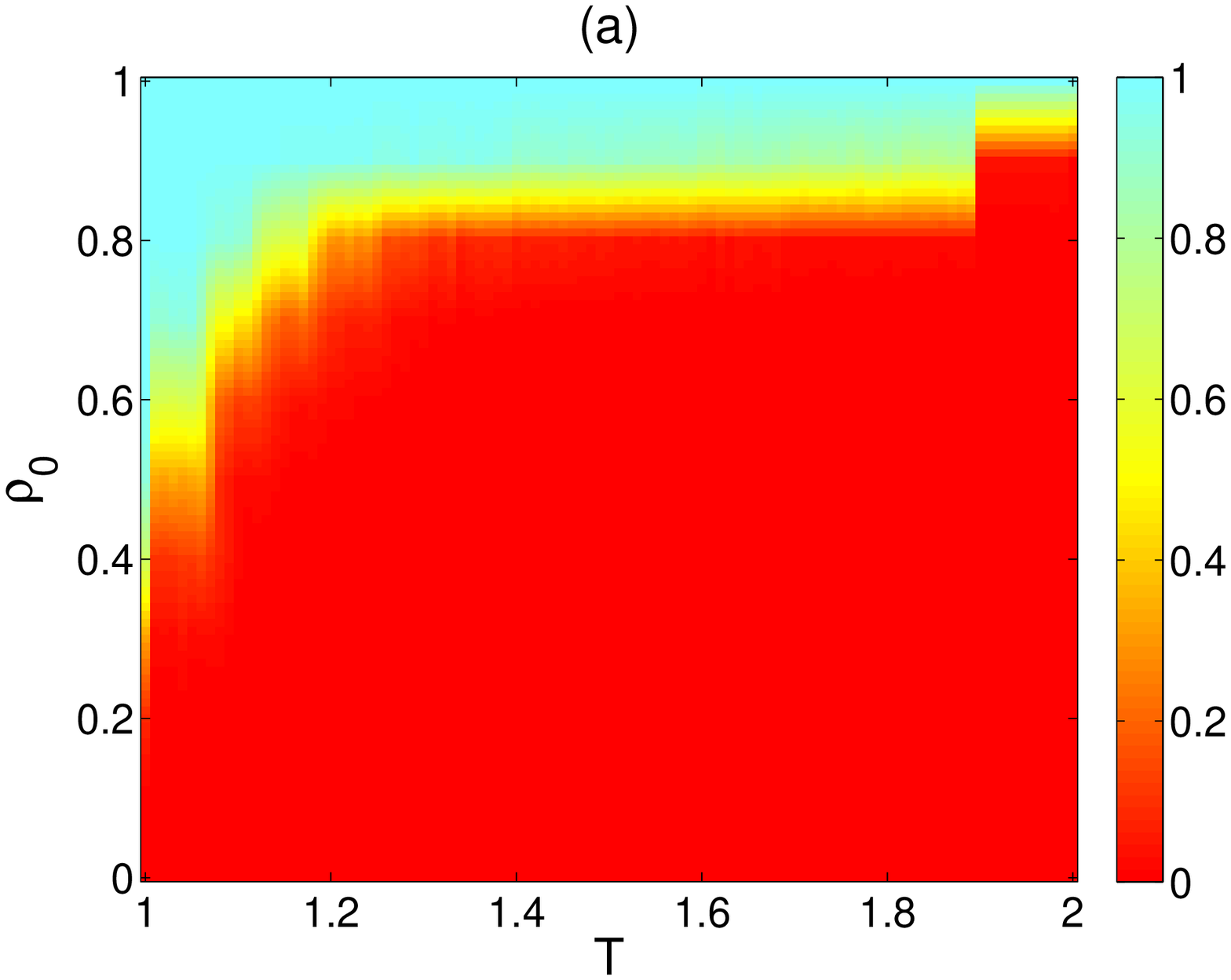}
\includegraphics[width=0.45\linewidth]{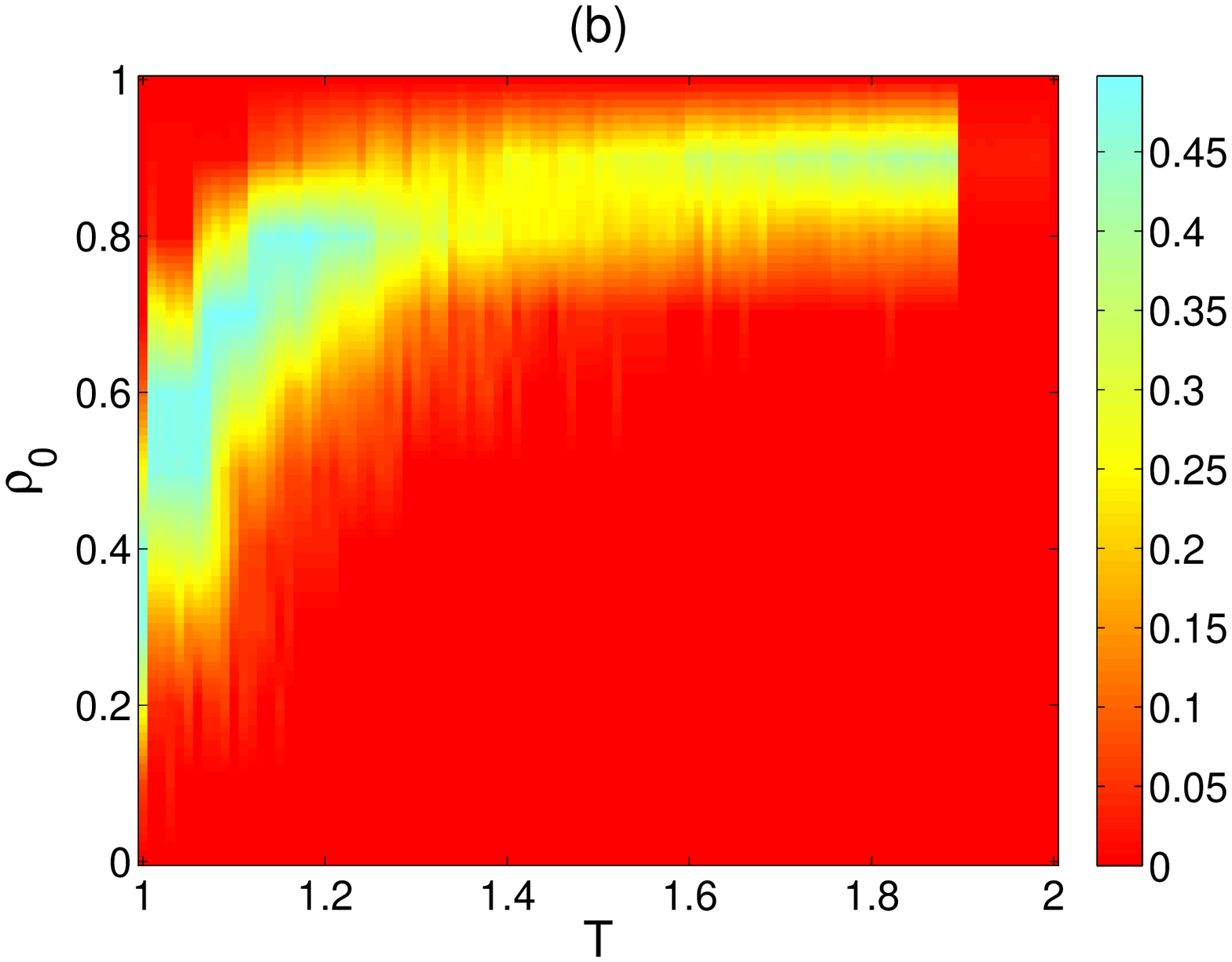} \\
\includegraphics[width=0.45\linewidth]{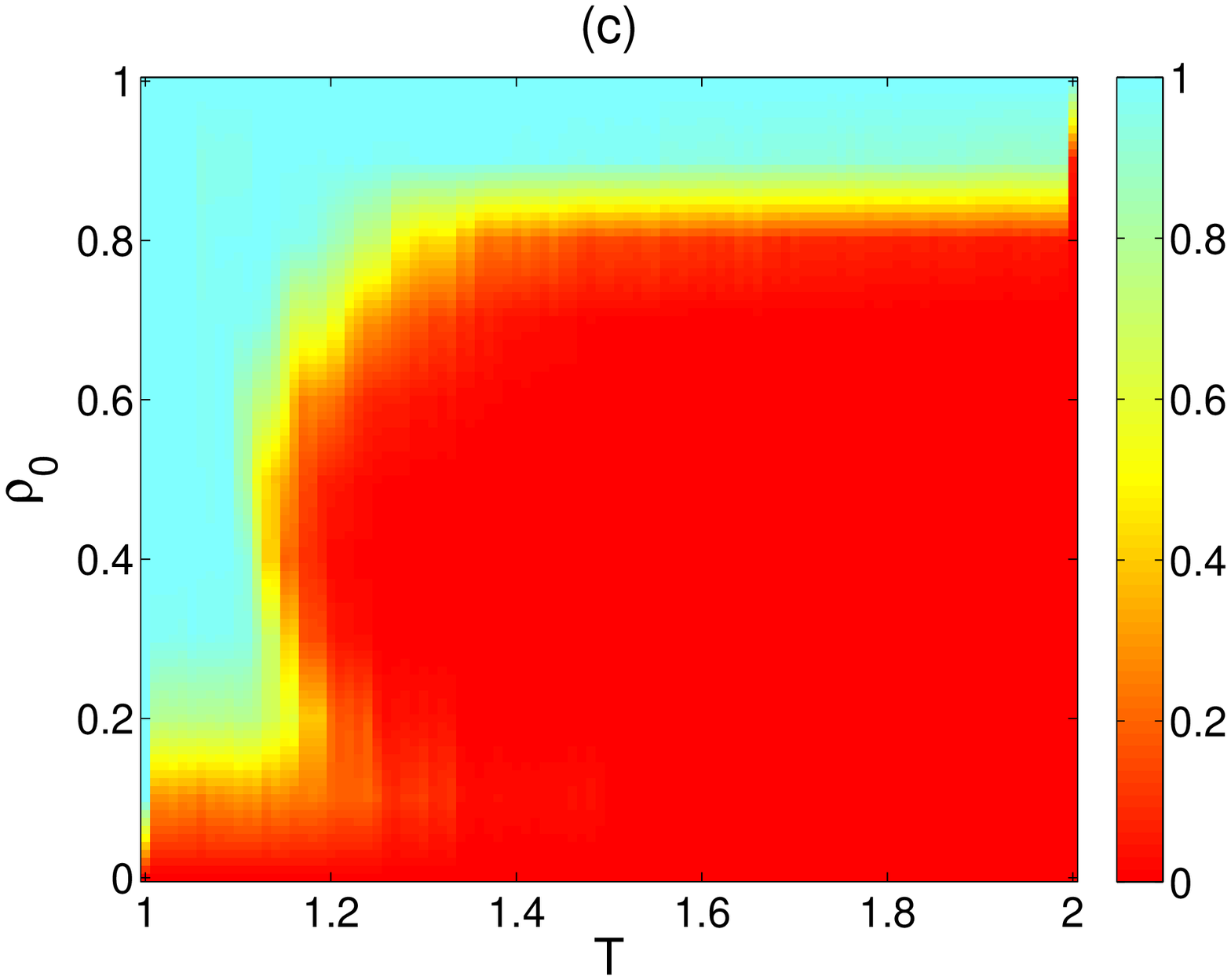}
\includegraphics[width=0.45\linewidth]{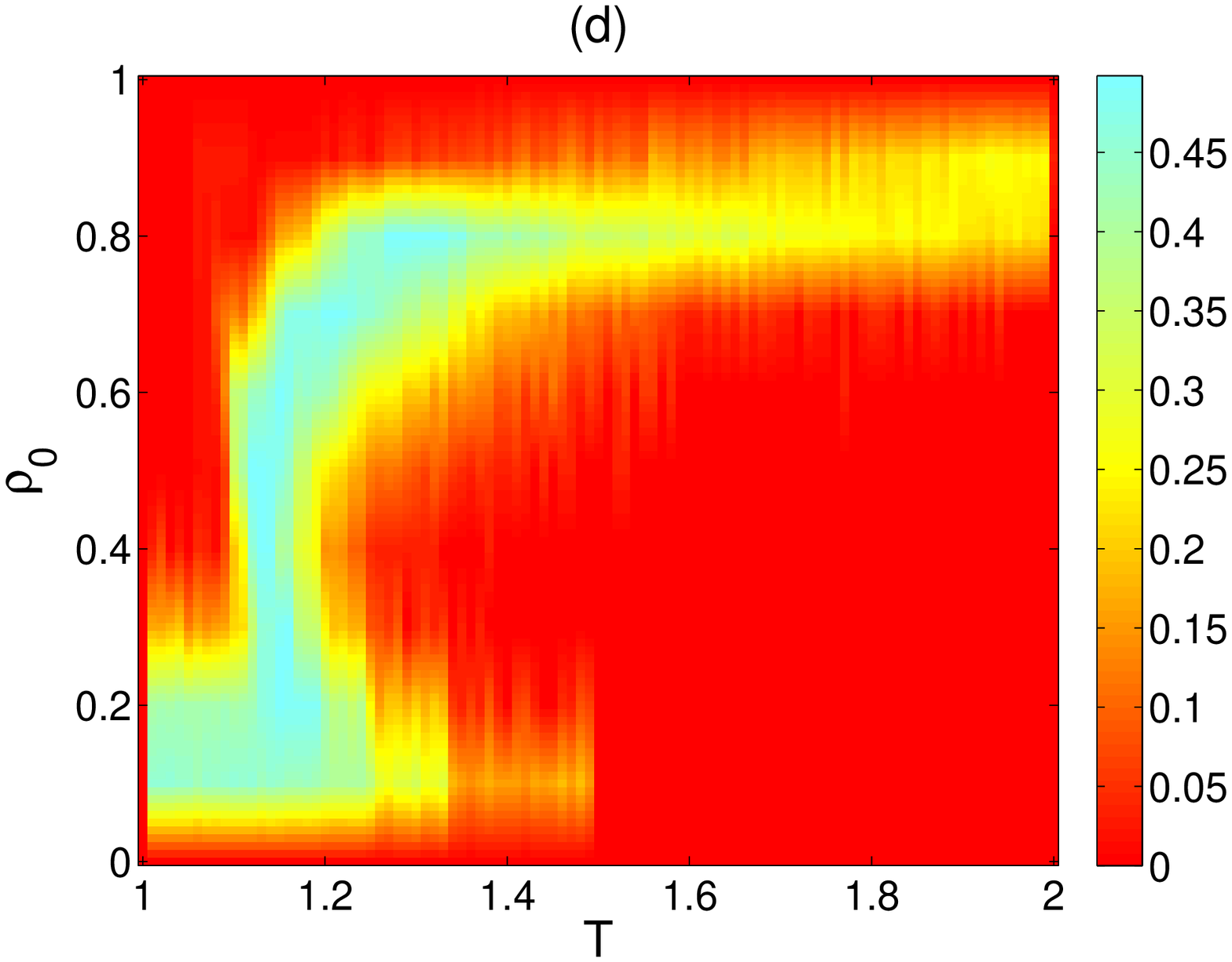}
\caption{
(a) Top view of Figure \ref{fig_superficies_converg}a, $\rho_\infty$ as function of $T$ and $\rho_0$ for $z = 20$ (without self-interaction);
(b) standard deviation of $\rho_\infty$ as function of $T$ and $\rho_0$ for $z = 20$ (without self-interaction);
(c) Top view of Figure \ref{fig_superficies_converg}b, $\rho_\infty$ as function of $T$ and $\rho_0$ for $z = 19$ (with self-interaction);
(d) standard deviation of $\rho_\infty$ as function of $T$ and $\rho_0$ for $z = 19$ (with self-interaction).
\label{fig_contornos_converg}}
\end{figure}

Figures \ref{fig_contornos_converg}a and \ref{fig_contornos_converg}c are the top view of the $\rho_\infty$ surfaces for $z = 20$ and $z = 19$, respectively.
They show the convergence of $\rho_\infty$ for even and odd $z$.
Figures \ref{fig_contornos_converg}b and \ref{fig_contornos_converg}d are the $\rho_\infty$ standard deviation for $z = 20$ and $z = 19$, respectively.
A relevant difference between even and odd $z$ is that the cooperative phase persists for $T < 1.1$ in the range of $0.1 < \rho_0 < 0.5$ in the presence of self-interaction (see Figure \ref{fig_contornos_converg}c).
If self-interaction is present the shift in $T_c$ to higher values remains in higher values of $z$.

\begin{figure}[!htb]
\includegraphics[width=0.45\linewidth]{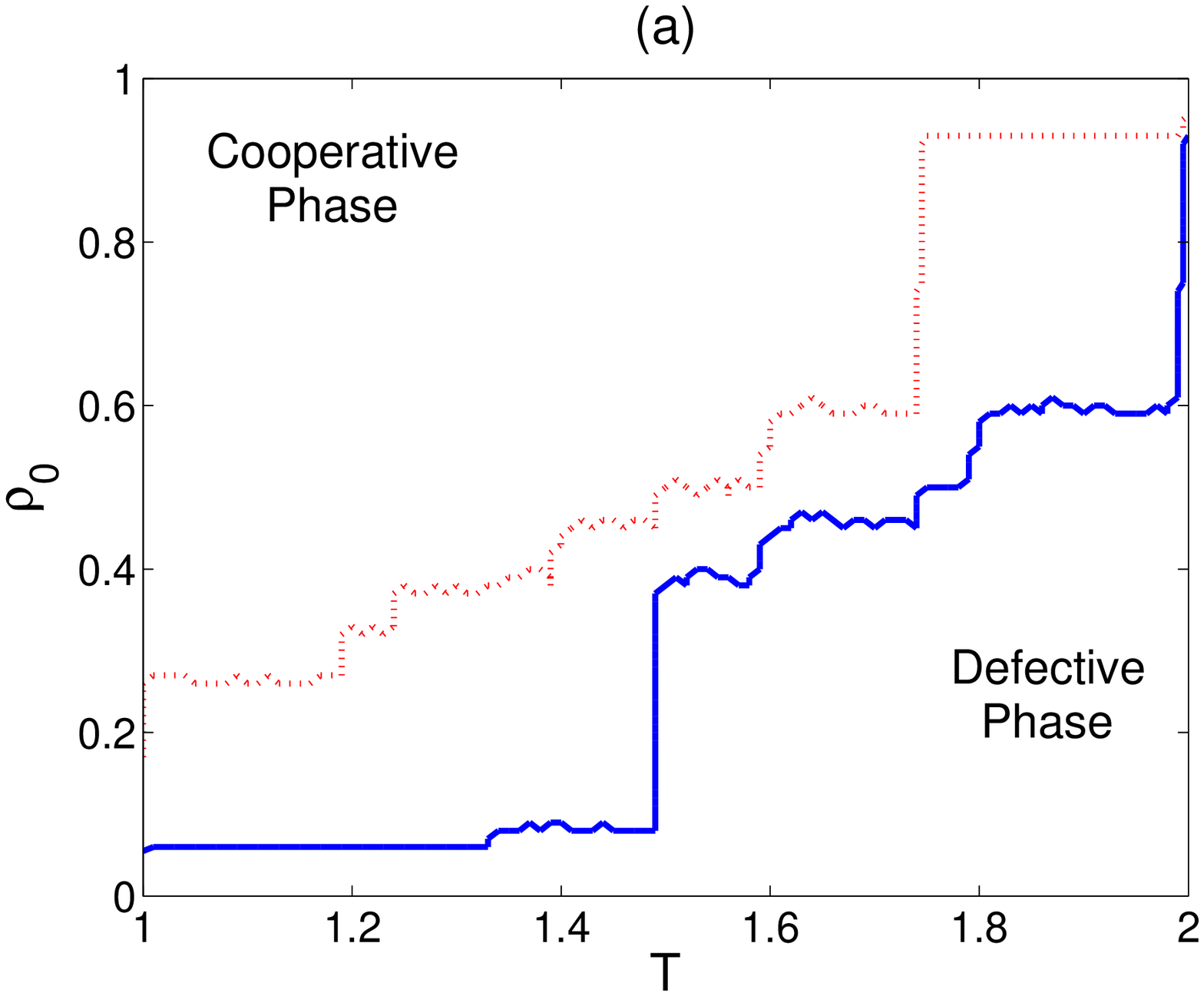}
\includegraphics[width=0.45\linewidth]{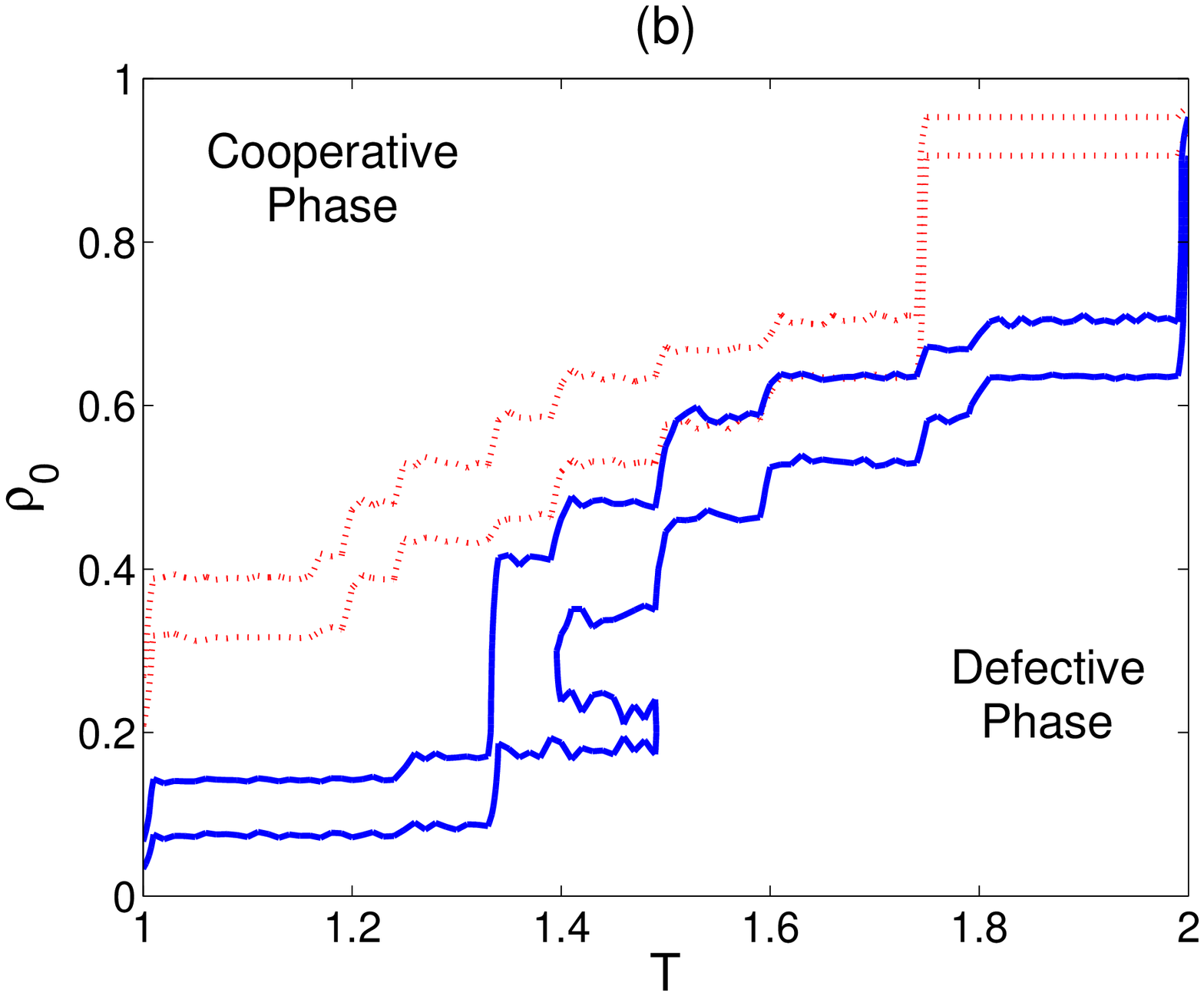}\\
\includegraphics[width=0.45\linewidth]{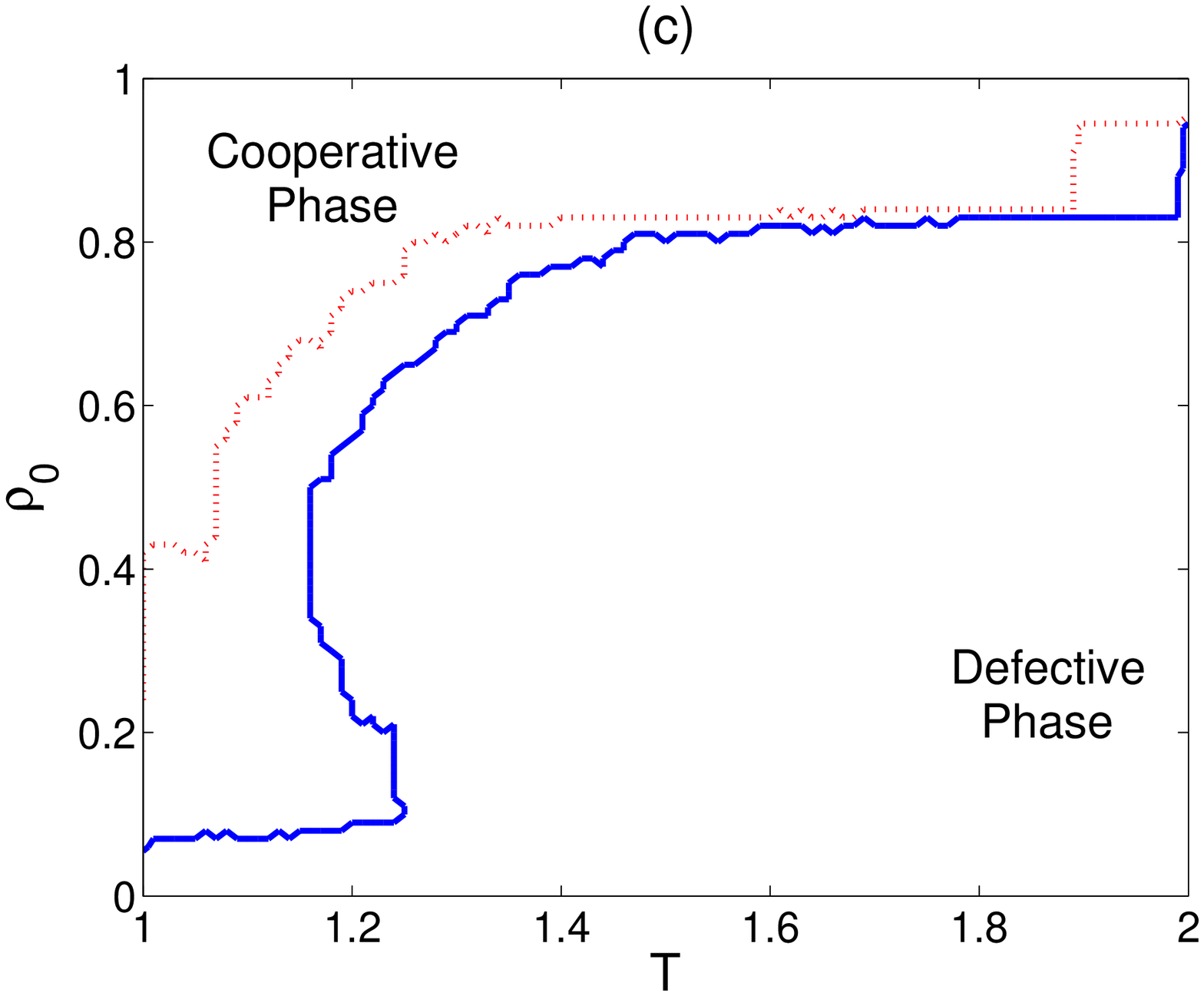}
\includegraphics[width=0.45\linewidth]{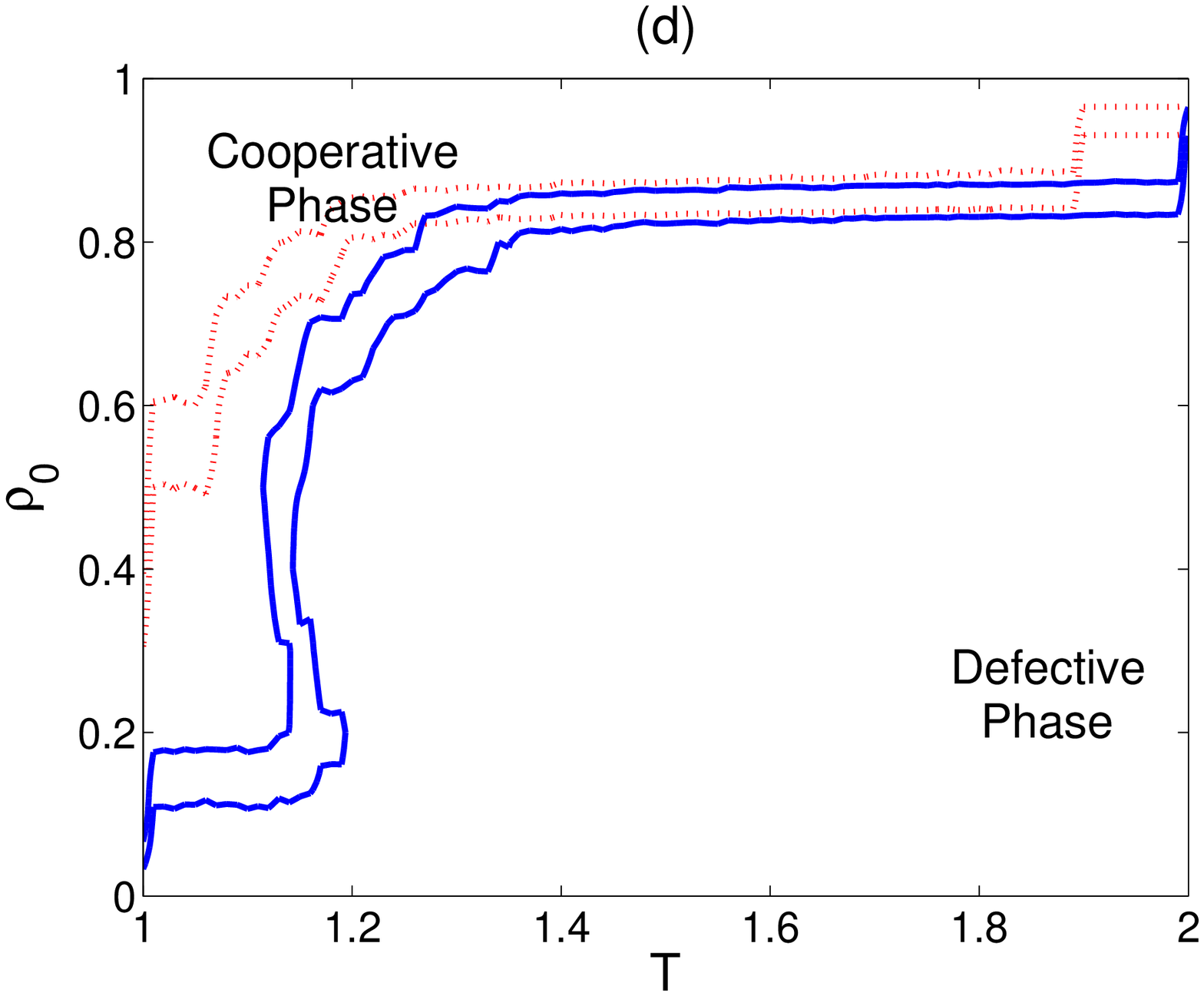}
\caption{
Phase diagram of $\rho_\infty$ in the parameter space.
(a) $z = 8$ and $z = 9$, contours of the cooperative/defective phase;
(b) $z = 8$ and $z = 9$, contours of the cooperative/coexistence/defective phase;
(c) $z = 20$ and $z = 19$, contours of the cooperative/defective phase;
(d) $z = 20$ and $z = 19$, contours of the cooperative/coexistence/defective phase.
The coexistence phase the phase is the space between the first and the second contours of the same $z$.
\label{fig_contornos_linha}}
\end{figure}

A slice of the plane $T \rho_0$ of the $\rho_\infty$ surface in Figures \ref{fig_superficies}a and \ref{fig_superficies}b at $\rho_\infty = 0.5$, may represent a phase diagram.
In Figures \ref{fig_contornos_linha}a and \ref{fig_contornos_linha}c, the contours separate the cooperative/defective phases, i.e. the phase-diagram.
Figures \ref{fig_contornos_linha}b and \ref{fig_contornos_linha}d are the contours that take into account the standard deviation.
Since there is the phase coexistence, these contours separate the cooperative/coexistence/defective phases.

\begin{figure}[!htb]
(a) \\
\includegraphics[width=1.0\linewidth]{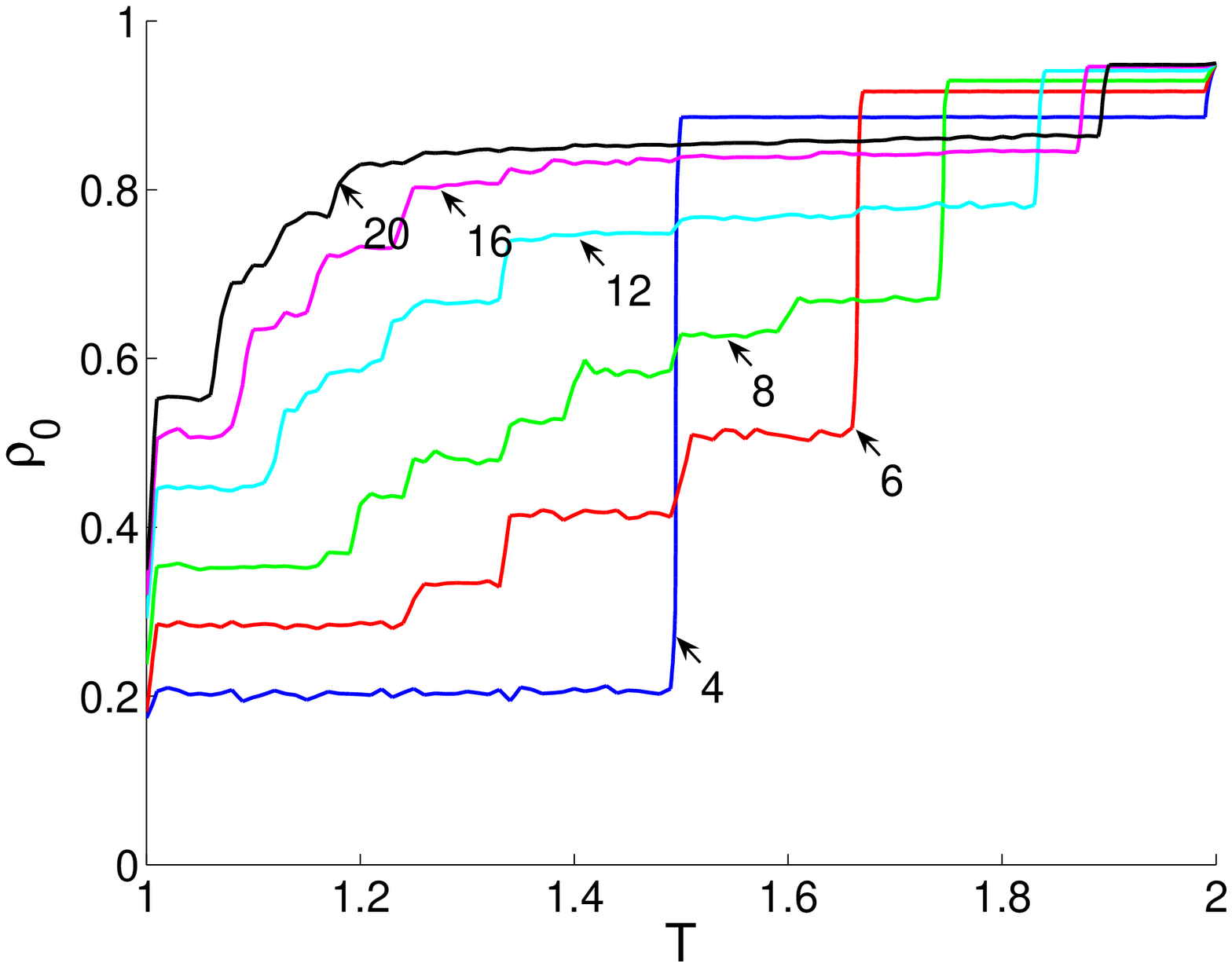} \\
(b) \\
\includegraphics[width=1.0\linewidth]{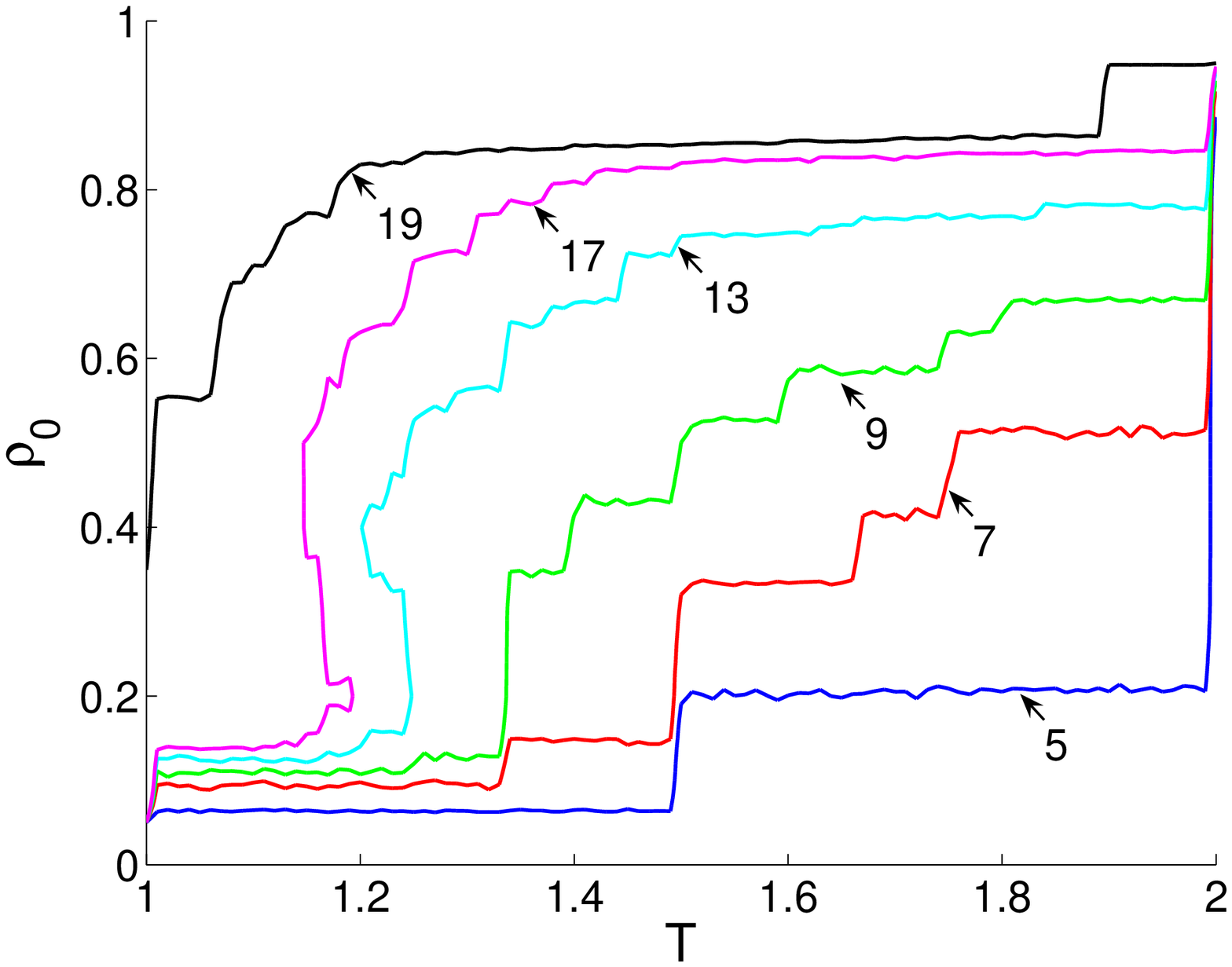}
\caption{
Phase diagram of $\rho_\infty$ in the parameter space.
(a) contours of the cooperative/defective phase for even $z = (4,6,8,12,16,20)$ (without self-interaction);
(b) contours of the cooperative/defective phase for odd $z = (5,7,9,13,17,19)$ (with self-interaction).
Notice the difference due to the parity of $z$ is not so important  for $z \gg 1$
 \label{fig_contornos_parimpar}}
\end{figure}

In Figures \ref{fig_contornos_parimpar}a and \ref{fig_contornos_parimpar}b, there are the contours of the cooperative/defective phase for different $z$ values.
When $z$ increases the contours converge to the same pattern independently if $z$ is even or odd as shown in Figure \ref{fig_contornos_linha}c and \ref{fig_contornos_linha}d.
For small $z$ values, the $z$ parity generates remarkable differences in the contours, if $z$ increases, the contours converge and present a similar form and the phase coexistence region is narrower than for small $z$ values.

\section{Conclusion} \label{conclusao}
The Prisoner's Dilemma in the one-dimensional cellular automata yields results according to the results obtained previously for regular lattices in $d$ dimensions.
The exhaustive exploration of the parameter space allows us to observe that the parameter $z$ plays the main role in the dynamics.
For low $z$ values, the influence of self-interaction is remarkable.
Some studies about the PD with variable coordination number, i.e. the neighborhood size $z$, have been carried out.
However, these studies adopt lattice topologies that are different from the one-dimensional lattice used here, e.g. square lattice \cite{ifti2004}, complex networks as random graphs \cite{duran2005}, scale-free networks \cite{ifti2004,santos2005}, small-world networks \cite{abramson2001}.
Another difference in comparison to these studies is that the state update of the players is asynchronous \cite{ifti2004,santos2005}, but in our case is synchronous.
Despite these differences, the main features due to the $z$ variation remain, such as the dependence on the asymptotic cooperator proportion on the neighborhood size.
Our results are similar to those obtained by Dur\'{a}n and Mulet \cite{duran2005} considering the neighboorhood with  self-interaction (odd $z$).
Comparing our results to those found in the literature, it is possible to see that the way the connection among the players is settled plays another important role in this problem independently of the space dimensionality or network structure.

For intermediate values of $T$ and $\rho_0$ the chaotic phase occurs.
In the chaotic phase the outcome $\rho_\infty$ can belong to the cooperative or defective phase due to only a small change in the initial distribution of the cooperators.

\section*{Acknowledgments}
M. A. P. would like to thank CAPES for the fellowship.
A. S. M. acknowledges the agencies CNPq (305527/2004-5) and FAPESP (2005/02408-0) for support.
A. L. E. would like to thank CNPq for the fellowship and FAPESP and MCT/CNPq Fundo Setorial de Infra-Estrutura (06/60333-0) for the financial support.

\end{document}